\pdfoutput=1
\documentclass{article}

\usepackage{algorithm2e}
\usepackage{algpseudocode}

\usepackage{epstopdf,graphicx,caption,subcaption,setspace}
\usepackage{multirow,amsmath,amssymb,url,amsthm}
\usepackage{amsfonts,amsmath,amssymb}
\usepackage{blindtext,stmaryrd}
\usepackage{multirow}

\usepackage{calc}%     needed for the width/height calculations

\newcommand{\eqdef}{\overset{c}{\equiv}}

\newtheorem{definition}{Definition}
\newtheorem{proposition}{Definition}

\begin{document}

\title{Privacy Preserving PageRank Algorithm By Using Secure Multi-Party Computation}
%

\author{Ferhat \"{O}zg\"{u}r \c{C}atak \vspace{10pt} \\T\"{U}B\.{I}TAK B\.{I}LGEM Cyber Security Institute, Kocaeli, Turkey\\ozgur.catak@tubitak.gov.tr}
\date{}
%


%
%\institute{T\"{U}B\.{I}TAK B\.{I}LGEM Cyber Security Institute, Kocaeli, Turkey,\\
%\email{ozgur.catak@tubitak.gov.tr}}

\maketitle              % typeset the title of the contribution

\begin{abstract}
In this work, we study the problem of privacy preserving computation on PageRank algorithm. The idea is to enforce the secure multi party computation of the algorithm iteratively using homomorphic encryption based on Paillier scheme. In the proposed PageRank computation, a user encrypt its own graph data using asymmetric encryption method, sends the data set into different parties in a privacy-preserving manner. Each party computes its own encrypted entity, but learns nothing about the data at other parties.
%\keywords{secure multi party computation, homomorphic encryption, privacy, PageRank}
\end{abstract}

\section{Introduction}
In this work, we investigate the problem with two or more parties wish to compute the PageRank algorithm \cite{Pageetal98} on an encrypted graph data set. In this computation model a user doesn't want to disclose private graph data and want to compute compute intensive large scale graph data set. 

Graph data sets often contain some private feature information about personal identifying information and also their sensitive relationships \cite{brickell2005privacy,zheleva2008preserving,zhou2008brief}.  The disclosure risk of the sensitive data arise with the outsourced computation such cloud computing based computation model. Secure multi party computation based privacy preserving techniques are usually adopted to privacy through secret sharing and homomorphic encryption scheme for the original graph data.

In this paper, we investigate how to perform the PageRank computation over the encrypted graph data. The main contributions are
\begin{itemize}
	\item The work protects the all structure of a graph data without losing the capability to perform PageRank algorithm over it.
	\item The proposed method is secure in the semi-honest model. Assuming that all parties correctly follow the protocol.
\end{itemize}

The rest of the paper is organized as follows. Section \ref{sec:privacy_definition} defines the privacy, secure multi-party computation and semi-honest security model. Section \ref{sec:preliminaries} introduces homomorphic encryption, floating point numbers problem and arbitrarily partitioned data. Section \ref{sec:experiments} presents our method to compute the PageRank values on the encrypted synthetic graph data. Section \ref{sec:conclusion} concludes the paper.

\section{Privacy definition}\label{sec:privacy_definition}
To decide whether a solution achieves the privacy requirements, we need to know the formal definition of privacy. We use a simplified form of the standard definition of security in the static, semi-honest model due to Goldreich \cite{goldreich1987play}. Our other assumption is that all parties follow the protocol which is called semi-honest security model \cite{lindell2000privacy,yang2005privacy}. A formal definition of private two-party computation in the semi-honest model is given below.
\begin{definition}(privacy w.r.t. semi-honest behavior): Let $f: \{0,1\}^* \times \{0,1\}^* \mapsto \{0,1\}^* \times \{0,1\}^*$ be probabilistic polynomial-time functionality, where $f_1(x,y)$ (respectively $f_2(x,y)$) denotes first (resp., second) element of $f(x,y)$; and let $\Pi$ be two-party protocol for computing $f$. The view of the first (resp., second) party during an execution of $\Pi$, denoted $view_1^\Pi$ (resp., $view_2^\Pi$), be ($x, r_1, m_1, \cdots, m_t$) (resp., ($y, r_2, m_1, \cdots, m_t$)). $r_1$ denotes the outcome of the first (resp., $r_2$ second) party's internal coin tosses, and $m_i$ denotes the $i^{th}$ message it has received.
	
The output of the first (resp., second) party during an execution of $\Pi$ on $(x,y)$, denoted $OUTPUT_1^{\Pi}(x,y)$ (resp., $OUTPUT_2^{\Pi}(x,y)$) is implicit in party's view of the execution.
	
We say that $\Pi$ privately computes $f$, if there exists polynomial-time algorithms, denoted $S_1$ and $S_2$, such that
$$\{(S_1(x,f_1(x,y)),f_2(x,y)\}_{x,y \in \{0,1\}^*} \eqdef \{VIEW_1^\Pi(x,y)\}_{x,y\in \{0,1\}^*}$$ 
$$\{(S_2(x,f_1(x,y)),f_2(x,y)\}_{x,y \in \{0,1\}^*} \eqdef \{VIEW_2^\Pi(x,y)\}_{x,y\in \{0,1\}^*}$$ 
\end{definition}

In this section formal definition of privacy is given in the graph data set that can model a variety of security protocols and attacks. The proposed model involves a graph $G(V,E)$, a set of $k$ parties $P_1, \cdots, P_k$. Given a matrix $D$, let $|D|$ denotes the number of rows in $D$.

We continue with the definitions for security in the semi-honest model. In semi-honest security model, Assuming that a party correctly follow the protocols using its correct input. 

\section{Preliminaries}\label{sec:preliminaries}
In this section, we briefly introduce preliminary knowledge of homomorphic encryption, floating point numbers and arbitrarily partitioned data. 
\subsection{Homomorphic Encryption}
Homomorphic encryption enables operations on plaintexts to be performed on their respective ciphertexts without disclosing the plaintexts when data is divided between two or more servers as it facilitates computations with ciphertexts. A public-key encryption scheme is additively homomorphic if, given two encrypted messages $Enc(a)$ and $Enc(b)$, there exists a public-key operation $\oplus$ such that $Enc(a) \oplus Enc(b)$ is an encryption of $a+b$. Formally, a cryptosystem is additively homomorphic if for any secret key, public key $(sk, pk)$ the plaintext space $\mathcal{P} = \mathbb{Z}_N$ for $x,y \in \mathbb{Z}_N$.

\begin{equation}
\begin{split}
E_{pk}(x + y \,\, mod \,\, N) & = E_{pk}(x).E_{pk}(y) \\
E_{pk}(x \cdot y \,\, mod \,\, N) & = E_{pk}(x)^y
\end{split}
\end{equation}
\subsubsection{Paillier's Encryption Scheme}
Paillier cryptosystem \cite{paillier1999public} is a probabilistic asymmetric algorithm based on the problem to decide whether a number is an $n$th residue modulo $n^2$ \cite{orlandi2007oblivious}. The problem of computing $n$th residue classes is believed to be computationally hard where $n$ is the product of two large primes.

Given a set of possible plaintexts $M$, a set of key pairs $K=PK \times SK$ where $PK$ is the public key, $SK$ is the secret key.
Paillier homomorphic encryption cryptosystem satisfies the following property of any two plaintexts $m_1$ and $m_2$ and a constant value $a$.
\begin{equation}
D_{sk} \left( E_{pk}\left(m_1\right) \times E_{pk}\left(m_1\right) \right) = m_1 + m_2
\end{equation}

\begin{equation}
D_{sk} \left( E_{pk}\left(m_1\right)^a \right) = a \times m_1
\end{equation}

\begin{proposition} If the adjacency matrix $A(G)$ of a graph $G$ has no loops, then all entries on the main diagonal of $A(G)$ are zeros. Hence $a_{ij}=0$ whenever $i=j \forall i,j$. 
\end{proposition}

The adjacency matrix $A(G)$ of a graph $G$ is sparse and often binary. In order to prevent guessing elements of input data set, Paillier cryptosystem has probabilistic encryption that does not encrypt two equal plain text with the same encryption key into the same ciphertext.	

\subsection{Floating Point Numbers}
Although the proposed protocol manipulate integers, the PageRank algorithm is typically applied to continuous data. However, in the case of real number inputs to the protocol, we need to map data vectors into the discrete domain \cite{kamm2015secure}. 

Let $ConvertInteger: \mathbb{R}^m \rightarrow \mathbb{Z}^m$ be the corresponding function that multiplies its argument by an exponent $(K: 2^K)$ then rounds them to the nearest integer value and thus support finite precision. Equation \ref{eq:convint} shows the conversion function.
\begin{equation}
\label{eq:convint}
\hat{\mathbf{x}} \gets ConvertInteger(\mathbf{x}) \,\,\, where \,\, \mathbf{x} \in \mathbb{R}^m, \hat{\mathbf{x}} \in \mathbb{Z}^m
\end{equation}

\subsection{Arbitrarily Partitioned Data}
In this work, arbitrary partitioned data between multi-parties ($K$), $K > 2$ is considered. In the arbitrary partitioned data scheme, there is no specific order of how the data is divided between multiple parties. Specifically, if we have a data set $X = \{\mathbf{x}_1, \cdots \mathbf{x}_n\}$, consisting of $n$ row, and each rows in $X$ contains $m$ numeric attributes $\mathbf{x}_i = \{x_i^1 \cdots x_i^m\}$. $X_i^j$ is the subset of data set owned by party $P_j$, then we have $X_i^1 \cup X_i^2 \cdots \cup X_i^K = X_i$ and $X_i^1 \cap X_i^2 \cdots \cap X_i^K = \emptyset $. In each row ($X_i$), party $P_k$ has a number of attributes $t_i^k$, where $\sum_{p=1}^{K}{t_i^k} = m$ and each party's attribute size does not have to be equal. If a party has the same attributes in each row, then the arbitrary partition becomes a vertical partition.

\section{Privacy-Preserving PageRank}
We now formally define the problem. Let $k$ be the number of parties, each having different attributes for the same set of entities (i.e. adjacency matrix rows). Parties $P_1, \cdots, P_k$ have as their respective private input sets (\textit{i.e.}, adjacency matrix) $S_1, \cdots, S_k$ drawn from some finite universe $U$.  The parties try to measure the importance of nodes with their joint data using the PageRank algorithm in a privacy-preserving manner, \textit{i.e.}, without leaking their elements of $S$.

\subsection{Notation}
A directed graph $\mathcal{G}(\mathcal{V},\mathcal{E},m)$  representing a web sites set is a set on $n$ nodes connected by a set of $m$ links, where $\mathcal{V}$ denotes the set of nodes and $\mathcal{E} \subseteq \mathcal{V} \times \mathcal{V}$ is the set of links and $m: \mathcal{V} \rightarrow (0, \infty)$ is a weight. For a node $u \in \mathcal{V}$, $\mathcal{N}$ denotes the number of hyperlinks contained in page (node) $u$.

Given $s_1, s_2 \in \mathcal{V}$, one can say that $x$ and $y$ are neighbors (\textit{i.e.}  hyperlinked from $x$ to $y$ ) if $\mathcal{E}(x,y) > 0$ \cite{baloudi2015adjacency}.

\subsection{Security model}
In this paper, the aim is to enable multiple parties (or servers) to jointly conduct the PageRank computation without revealing their private data. Our main assumption is that the input data set is divided between two or more parties, that are willing to compute the PageRank of $\mathcal{G}$ if nothing beyond the expected end results are revealed \cite{bansal2011privacy}. Our other assumption is that all parties follow the protocol which is called semi-honest security model \cite{lindell2000privacy,yang2005privacy}.

\subsection{Graph Representation}
In order to protect the unauthorized access to the sensitive graph information, one needs to represent and encrypt the graph in a proper way. In order to achieve these requirements, adjacency matrix representation is used. let $k$ be the number of parties, each has the same attributes for the different entities. The distributed parties try to calculate PageRank algorithm by using their joint data (\textit{i.e.}, adjacency matrix). 

The PageRank algorithm needs adjacency matrix $\mathcal{A}$ of graph $\mathcal{G}(\mathcal{V},\mathcal{E})$. The proposed protocols directly follow the standard the PageRank algorithm. The approximations to the page rank values are iteratively computed until the changes in values in one iteration is below a threshold or max iteration number is reached. At each iteration, each party multiplies its own encrypted adjacency matrix $\llbracket \mathcal{A} \rrbracket_i$ with \textit{PageRank} and \textit{out degree} vectors as shown in Equation \ref{eq:pr_t}. 

\begin{equation}
   \llbracket \mathcal{T} \rrbracket_i = \begin{bmatrix} \llbracket a \rrbracket_{1,1} & \cdots & \llbracket a \rrbracket_{1,j} \\ \vdots& \ddots & \vdots \\ \llbracket a \rrbracket_{m,1} & \cdots & \llbracket a \rrbracket_{m,j} \end{bmatrix} \begin{bmatrix} Pr_1 \\ \vdots \\ Pr_m \end{bmatrix} \begin{bmatrix} \frac{1}{d_1} \\ \vdots \\ \frac{1}{d_m} \end{bmatrix} 
   \label{eq:pr_t}
\end{equation}

\noindent where $ \llbracket a \rrbracket$ is the encrypted elements of $ \llbracket \mathcal{A} \rrbracket$, $Pr$ is the PageRank values of graph $\mathcal{G}$ and $d$ is the out degree values. Once the $\llbracket \mathcal{T} \rrbracket_i$ are computed at each party, then another trusted party merges the matrices to find a global encrypted $\llbracket \mathcal{T} \rrbracket$ matrix as shown in Equation \ref{eq:t_union}.

\begin{equation}
	\llbracket \mathcal{T} \rrbracket = \bigcup _{i=1}^k \llbracket \mathcal{T} \rrbracket_i
	\label{eq:t_union}
\end{equation}

The trusted party decrypts $\llbracket \mathcal{T} \rrbracket$ using private key $key_{priv}$, calculates the PageRank values as shown in Equation \ref{eq:pr}.
\begin{equation}
	Pr(\mathcal{V}_i) = \frac{1-d}{\mathcal{N}} + d\sum_{p_j \in M(p_i)}\frac{Pr(\mathcal{V}_j)}{L(\mathcal{V}_j)}
	\label{eq:pr}
\end{equation}

\begin{algorithm}[t]
	\SetAlgorithmName{Protocol}{}
	\KwData{\textbf{Input:} Adjacency matrix: $\mathcal{A} \in \mathbb{R}^{m \times m}$ party size: $k$, Crypto key length: $l$, damping factor $d$, \\ }
	\KwResult{Encrypted sub data sets for each party $P_s$}
	\Begin{
		$\mathcal{D} \longleftarrow normalize(\mathcal{D})$\;
		$Key_{pub}, Key_{priv} \longleftarrow KeyGen(l)$ \Comment{Generate public/private keys} \;
		\For{$i\in k$}{
			// {\scriptsize \textit{create sub adjacency matrix $\mathcal{A}_i$ with random feature index for party $P_i$}} \;
			$\llbracket \mathcal{A}_i \rrbracket \longleftarrow encrypt(\mathcal{A}^i, Key_{pub})$\;
			$sendToParty(\llbracket \mathcal{A} \rrbracket_i, Key_{pub})$\;
		}
		\Repeat{Pagerank converges}{
			\ForEach{$P_s \in P$}{
				call \textit{secureIntMatrix( $\llbracket \mathcal{A} \rrbracket_s$)} \{Protocol \ref{alg:compute}\} \Comment{for each party}\;
			}
			$comp\_mat \gets \bigcup _{i=1}^kdecrypt( \llbracket \mathcal{T} \rrbracket_i,key_{priv})$ \Comment{Decrypt, merge}\;
			\For{$i = 1 \cdots m$}{
				$PageRank[i] \gets \frac{1-d}{N} + d \sum_{j = 1}^{m}comp\_mat[j,i]$
			}	
		}
		
	}
	\caption{Adjacency matrix split and encryption \label{alg:init}}
\end{algorithm}

\begin{algorithm}[t]
	\SetAlgorithmName{Protocol}{}
	\KwIn{\textbf{Input:} Encrypted sub adjacency matrix: $\llbracket \mathcal{A} \rrbracket_s \in \mathbb{R}^{m \times n}$, Public crypto key: $Key_{pub}$, scaling factor $c$\\ }
	\KwResult{Intermediate results of $computation\_matrix_s$}
	\Begin{
		$\llbracket comp\_mat\rrbracket_s = zeros(m,n)$ \;
		\For{$i = 1 \cdots n$}{
			$\mathbf{col} \gets \llbracket \mathcal{A} \rrbracket_s[:,i]$ \Comment{Get $i^{th}$ column vector of $\llbracket \mathcal{A} \rrbracket_s$ }\;
			\For{$j = 1 \cdots m$ }{
				$\llbracket comp\_mat\rrbracket_s[i,j] = \llbracket \mathcal{A} \rrbracket_s[i,j] \otimes \left( \frac{PageRank[i]}{OutDegree[i]} \cdot 10^c \right)$ \Comment{Encrypted multiplication} \;
			}
		}
		\textbf{return} $\llbracket comp\_mat\rrbracket_s$
	}
	\caption{secureIntMatrix \label{alg:compute}}
\end{algorithm}

At each iteration, the PageRank is calculated, \textit{i.e.}, the proposed protocols securely find the PageRank vector for graph $\mathcal{G}$. Once the initial PageRank values are known, the next Pagerank values can be computed locally. Protocol \ref{alg:init} use Protocol \ref{alg:compute} (\textit{secureIntMatrix}) to compute secure intermediate matrix. 

The proposed algorithm is shown in Protocol \ref{alg:init} - \ref{alg:compute}.
\section{Experiments}\label{sec:experiments}
We evaluated the proposed algorithm on a synthetic graph data set with 20 nodes. We implemented the protocols in Python 2.7 with the Paillier \footnote{https://github.com/mikeivanov/paillier} library. The experiments were performed on a computer with a 2.6 GHz Intel Core i5 processor and 4 GB main memory running Mac OSX. The execution time is measured by used processor time in seconds with different node size and different encryption key length. Experimental results are shwon in Table \ref{tbl:results}.
\begin{table}[t!]
   \caption{Experimental results}
   \label{tbl:results}
	\begin{center}
	\begin{tabular}{l l l l l}
	\hline\noalign{\smallskip}
	Party Size & 128 bit & 256 bit & 512 bit & 1024 bit\\
	\noalign{\smallskip}
    \hline
    \noalign{\smallskip}
    3   & 56.45 & 107.13 & 678.29 & 3711,48\\
    5   & 57.14 & 110.14 & 685.70 & 3506,84\\
    7   & 56.94 & 106.70 & 703.70 & 3416,14\\    
    10 & 53.65 & 108.72 & 962.98 & 3427,22\\
    \hline
	\end{tabular}
	\end{center}
\end{table}

\section{Conclusion and Future Works}\label{sec:conclusion}
We introduce a new homomorphic encryption based privacy-preserving PageRank algorithm with secure multi-party computation approach. We showed the effects of different encryption key length  on results with graphics. In particular, different party size and key length could also be investigated for both in computation time and computation error because of scaling problem of the encryption method.

\bibliographystyle{ieeetr}
\bibliography{references}

\end{document}